# Entropy driven reverse-metal-to-insulator transition and delta-temperatural transports in metastable perovskites of correlated rare-earth nickelate

*Jikun Chen[1], Haiyang Hu[1], Takeaki Yajima[2], Jiaou Wang[3], Binghui Ge[4], Hongliang Dong[5], Yong Jiang[1], Nuofu Chen[6], et al*

[1]Beijing Advanced Innovation Center for Materials Genome Engineering, School of Materials Science and Engineering, University of Science and Technology Beijing, Beijing 100083, China

[2]School of Engineering, The University of Tokyo, 2-11-16 Yayoi, Bunkyo-ku, Tokyo 113-0032, Japan

[3]Beijing Synchrotron Radiation Facility, Institute of High Energy Physics, Chinese Academy of Sciences, Beijing 100049, China

[4]Institute of Physical Science and Information Technology, Anhui University, 230601, Heifei, Anhui, China

Cleveland, Ohio 44106, United States

[5]Center for High Pressure Science and Technology Advanced Research, Shanghai 201203, China

[6]School of Renewable Energy, North China Electric Power University, Beijing 102206, China

Correspondence and request for materials: Prof. Jikun Chen (jikunchen@ustb.edu.cn). The present manuscript is in discussion with several external expert and the author number may increase when performing final submission.

**Abstract:**

   The metal to insulator transition (MIT) in Mott-Hubbard systems is one of the most important discoveries in condensed matter physics, and results in abrupt orbital transitions from the insulating to metallic phases by elevating temperature across a critical point ($T_{MIT}$). Although the MIT was previously expected to be mainly driven by the orbital Coulomb repulsion energy, the entropy contribution to the orbital free energy that also determines the relative stability of the metallic and insulating phases was largely overlooked. Herein, we demonstrate an orbital-entropy dominated reversible electronic phase transition in the metastable perovskite family of correlated rare-earth nicklates ($Re\text{NiO}_3$), in addition to their previously known MIT driven by orbital Coulomb energies. In reverse to MIT, the resistivity of $Re\text{NiO}_3$ abruptly increases by 2-3 orders by elevating $T$ across another critical point ($T_{R\text{-}MIT}$) below $T_{MIT}$, and such transition is named as reverse-metal to insulator transition (R-MIT). Combining the afterwards exponentially decreasing resistivity in the insulating phase of $Re\text{NiO}_3$ at further temperature elevation, a distinguished delta-temperatural transport character is established, which is potentially applicable for locking the working temperatures range for electric devices. The $T_{R\text{-}MIT}$ is shown to be enhanced via reducing the compositional complexity and size of $Re$ or imparting bi-axial compressive strains, and meanwhile the transition sharpness of delta-temperatural transport is reduced. Our discovery indicates that temperature range for a thermodynamically stable insulating phase of $Re\text{NiO}_3$ is in between of $T_{R\text{-}MIT}$ and $T_{MIT}$, while a new conductive phase with high orbital entropy is formed by further descending temperature below $T_{R\text{-}MIT}$.

The discovery of new electronic phases and reversible orbital transitions among multiple-states within electron correlated materials can enrich distinguished material functionalities beyond conventional and promote new applications [1-6]. The past century witnessed the development of the metal to insulator transition (MIT) discovered in the $d$-band correlated systems [7-15] that promotes new applications such as thermochromism [10], thermistor [11], electronic field effect transistors [12-14], and neuron-spin logical devices [15]. During MIT, the orbital configurations within Mott-Hubbard systems experience a sharp transition at a critical temperature ($T_{MIT}$) that abruptly switches the material phase between metal and insulator (semiconductor) [7-9]. In previous investigations, the MIT mainly attributed to the reduction in orbital Coulomb energy [4-15], while the respective variations in orbital configuration entropy ($S_{Orbit}$) were less considered as a significant contribution to the orbital Gibb's free energy ($\Delta G_{Orbit} = \Delta U_{Coul.} - T\Delta S_{Orbit}$). It is worth noticing that the elevation in phonon entropy should be also an important contribution to the MIT of vanadium dioxides when transits from the low symmetry monoclinic-insulating phase to the high symmetry rutile- metallic phase, as previously pointed out in ref [16]. Recently, several reports in high entropy compounds highlight the importance of the large elevation in the configuration entropy to the stability of material phases stability, which exceeds the contribution from the enthalpy to the free energy [17]. Beyond reasonable doubts, the synthesizing temperature can be effectively reduced for structural ceramics via introducing the compositional complexity or enlarging the configurational disorder. Extending an analogical consideration to electron correlated system with complex electronic phase diagram, it sheds a light on alternative manipulations on the relative stability among the multiple electronic phases ($\Delta G_{Orbit}$) from the aspect of $S_{Orbit}$, instead of $U_{Coul.}$, to achieve a new reversible phase transition beyond MIT.

It is interesting to note the temperature ($T$) induced variations in $S_{Orbit}$ for the metastable perovskite family of rare-earth nickelates ($Re$NiO$_3$), which are typical Mott-Hubbard MIT correlated system and exhibit continuously adjustable $T_{MIT}$ within 100-600 K via their rare-earth composition [8]. By reducing $T$ from their conventional metallic phase across $T_{MIT}$, the $Re$NiO$_3$ experience symmetry breaks form the aspects of both structures (via Jahn-Teller distortion) and orbital degeneracy (via charge disproportionations) [18-20]. Although the reduced symmetry is expected to abruptly decrease the $S_{Orbit}$ ($\Delta S_{Orbit,M \rightarrow I} < 0$), its positive contribution ($-T\Delta S_{Orbit,M \rightarrow I} > 0$) to $\Delta G_{orbit.}$ is smaller compared to the reduction in $U_{Coul.}$ ($\Delta U_{Coul.M \rightarrow I} < 0$) that dominates the transition from metallic to insulating phase. Further cooling down is expected to continuously descend $S_{Orbit}$ via improving the orbital ordering for the disproportionated $t^6_{2g}e^0_g$ (Ni$^{2+}$) and $t^6_{2g}e^1_g$ (Ni$^{4+}$) within the insulating phase of $Re$NiO$_3$ that gradually opens up the electronic band gap. This is evidenced by the thermistor transportations with large magnitudes of negative temperature coefficient of resistance (NTCR) beyond conventional semiconductors, as previously observed extensively in the insulating phase of several $Re$NiO$_3$ [11]. Noticing that this process is more dominated by the variation in the orbital ordering rather than the orbital configuration energy, a more significant temperature induced variations in $S_{Orbit}$ is expected, compared to the one in $U_{Coul.}$. If there is another critical temperature, at which the magnitude in the positive contribution to $\Delta G_{Orbit}$ from the entropy aspect ($-T\Delta S_{Orbit.}>0$) can completely offset the reduction in $U_{Coul.,M\rightarrow I}$, it will draw the potential to transit $Re$NiO$_3$ towards a new electronic phase with high $S_{Orbit}$ and low $U_{Coul.}$.

In this work, we demonstrate the presence of an orbital entropy dominated reversible electronic phase transition in the correlated perovskite family of rare-earth nicklates with high thermodynamical metastabilities ($Re$NiO$_3$: size of $Re$ < Nd), in addition to their MIT driven by orbital Coulomb energies. In such a transition, the resistivity of $Re$NiO$_3$ abruptly increases by 2-3 orders when elevating $T$ across another critical point ($T_{R-MIT}$) below $T_{MIT}$, the transportation of which is in reverse to MIT and is named as reverse-metal to insulator transition (R-MIT). This discovery indicates that temperature range for a thermodynamically stable insulating phase of $Re$NiO$_3$ should be in between of $T_{R-MIT}$ and $T_{MIT}$, while a new conductive phase with high orbital entropy is formed by descending the temperature below $T_{R-MIT}$. Noticing that the resistivity of $Re$NiO$_3$ firstly increases abruptly when elevating $T$ across $T_{MIT}$ and afterwards exponentially decreases via its reported NTCR thermistor transportations, it establishes a distinguished character of delta-temperatural transports. By adjusting the rare-earth compositions and imparting interfacial strains, the delta-temperatural transports of $Re$NiO$_3$ can be further regulated, and this new functionality is expected to be used in to sense and lock the working temperatures of electronic devices.

To trigger as-proposed R-MIT driven by the contribution from $S_{orbit.}$, both the thermodynamic and kinetic aspects need to be taken into account. Thermodynamically, it requires the insulating phase to exhibit more significant variations in $S_{orbit.}$ compared to the ones in $U_{Coul.}$, when reducing the temperature. In that case, it is possible that the contribution from descending $S_{Orbit}$ to the $G_{Orbit.}$ of the insulating phase gradually offsets the lower $U_{Coul.}$ of its insulating phase compared to the metallic phase, as illustrated in Figure 1a. This is expected to be the feature for the insulating phase of $Re$NiO$_3$, noticing that their NTCR transportation is expected to be more associated to the variation in orbital ordering rather than the orbital potentials. Kinetically, the intrinsic metastability of $Re$NiO$_3$ elevates the free energies for both its metallic and insulating phases, the effect of which reduces the energy barrier prohibiting the occurrence of R-MIT at $T_{R-MIT}$, as illustrated in Figure 1b. In Figure 1c, the expected orbital transitions across $T_{MIT}$ and $T_{R-MIT}$ are illustrated. The metal to insulator transition via MIT when elevating $T$ across $T_{MIT}$ reduces $U_{Coul.}$ and increases $S_{Orbit}$, while the one via R-MIT when reducing $T$ across $T_{R-MIT}$ is expected to more significantly enhance $S_{Orbit}$ compared to $U_{Coul.}$.

As a typical example, Figure 2a shows the temperatural dependent resistivities ($R$-$T$) for quasi-single crystalline SmNiO$_3$ thin films grown on single crystalline perovskite substrates, such as LaAlO$_3$, SrTiO$_3$ and (La,Sr)(Al,Ta)O$_3$, using the chemical approach we described previously [11]. We can clearly observe as-proposed delta temperatural transportation behaviors for all three samples, in which cases the resistivity exponentially increases by reducing $T$ until $T_{R-MIT}$ (the delta-transition point) and afterwards abruptly reduces. The $R$-$T$ tendencies measured via heating up or cooling down overlaps with each other, indicating that both the R-MIT and its resultant delta-temperature transports are reversible.

It is also worth noticing that a higher $T_{Delta}$ is observed for $SmNiO_3/LaAlO_3$, compared to the ones for $SmNiO_3/SrTiO_3$ and $SmNiO_3/(La,Sr)(Al,Ta)O_3$. According to our previous reports [11], the lattice mismatches between the film and substrate results in various status of interfacial coherency and strains. As their cross-section interfacial morphology demonstrated in Figure 2b, the $SmNiO_3$ film is coherently grown on the $LaAlO_3$ substrate owing to a small lattice mismatch (~ 0.4%), and is under biaxial compressive interfacial strain. This is in contrast to $SmNiO_3/SrTiO_3$ or $SmNiO_3/(La,Sr)(Al,Ta)O_3$, in which cases the epitaxial coherency is not perservable owning to a large lattice mismatch (~ -2.4% and -1.6%), and thereby the interfacial strain in relaxed. The X-ray reciprocal space mapping (RSM) results for $SmNiO_3$ grown on the various substrates are further shown in Figure S1, where the same in-plane lattice vector is observed for $SmNiO_3$ and $LaAlO_3$. In contrast, the in-plane lattice vector for $SmNiO_3$ slightly differs to the $(La,Sr)(Al,Ta)O_3$ that indicates a relaxation in the tensile distortion, while the relaxation is more significant for $SmNiO_3/SrTiO_3$. The biaxial compressive distortion is known to reduce the $T_{MIT}$ of $SmNiO_3$, which is also observed in this work as shown in Figure S2. To further investigate their electronic structures, the near edge X-ray absorption fine structure (NEXAF) analysis was performed to probe the relative variations in the Ni: $L$-edge and O: $K$-edge of our samples, as shown in Figure 2c and 2d, respectively. Compared to the strain relaxed $SmNiO_3/SrTiO_3$ or $SmNiO_3/(La,Sr)(Al,Ta)O_3$, a larger proportion of the B proportion within the Ni: $L_3$ spectrum is observed for the compressively distorted $SmNiO_3/LaAlO_3$. This indicates the elevation in the proportion of the $t^6_{2g}e^1_g$ ($Ni^{3+}$) ground state orbital configuration compared to $t^6_{2g}e^2_g$ ($Ni^{2+}$) [21,22] when imparting the compressive distortion. A consistent variation was further observed in their O: $K$-edge (Figure 2d), in which case the pre-peak ($d^8L$) for $SmNiO_3$/LAO exhibits a higher intensity [21,23].

From the above results, we can see that imparting bi-axial compressive distortion upon $SmNiO_3$ elevates the $T_{R-MIT}$ during R-MIT, which is not simply associated to the manipulation in relative electronic phase stability in MIT as the $T_{MIT}$ is reduced. The situation for biaxial tensile distorted $SmNiO_3$, i.e. by coherently grown on $SrTiO_3$ via pulsed laser deposition similar to ref [24], is demonstrated in Figure S3, in which case the tensile distortion can be stabilized via the kinetics of the plasma involved process [24,25]. The same in-plane lattice vector is observed for the pulsed laser deposited $SmNiO_3$ and the $SrTiO_3$ substrate underneath (see Figure S3a), while a coherent interface is observed in their between (see Figure S3b). Nevertheless, no R-MIT behavior is observed for the biaxial tensile strained $SmNiO_3$ in the investigated temperature down to 2 K, indicating that the $T_{R-MIT}$ is either eliminated or further descended below 2K. This observation in R-MIT is analogies to the MIT of tensile strained $SmNiO_3$, in which case its MIT was not clearly observed [24,25].

To further regulate $T_{R-MIT}$ in a broader range of temperature, we adjusted the rare-earth composition occupying the A-site of the perovskite structure, similar to the regulations in their $T_{MIT}$ [8,26,27]. Reducing size of the rare-earth element results in more tilted NiO$_6$ octahedron that elevates the metastablility in their more distorted perovskite structure, as demonstrated in Figure 3a. This was previously known to strengthen the insulating phase at high temperature via more effectively split an energy gap within the hybridized O2$p$-Ni3$d$ orbits that elevates the $T_{MIT}$ [8]. It is also worth noticing that the enhanced structural distortion for using smaller $Re$ meanwhile reduces the symmetry in orbital configurations and is expected to reduce the initial $S_{Orbit}$. This provides a larger thermodynamical potential to trigger the R-MIT at a higher temperature, while the enhanced metastability is expected to kinetically reduce the transition energy barrier.

Our expectation is confirmed by comparing the $R$-$T$ results measured for $Re$NiO$_3$ with various single elemental $Re$ compositions grown on LaAlO$_3$ substrate, as shown in Figure 3b. By reducing size of $Re$ from Sm towards Tm, the temperature to trigger the R-MIT is observed to be elevated from ～50 K to ～150 K, while their $T_{R\text{-}MIT}$ is more clearly compared in Figure 3c. Meanwhile, the sharpness in the delta-temperatural transport is reduced, as indicated by their smaller variations in resistivity ($R_{T\text{-}Delta}/R_{300K}$) and the broadening of the full width half maximum of the delta-temperature range ($T_{Delta}$-FWHM) shown in Figure 3d and 3e, respectively. It is also interesting to note that as-achieved delta-temperatural transports is relatively stable in magnetic field up to 10 T, as demonstrated in Figure 3f for TmNiO$_3$/LaAlO$_3$ (see more example in Figure S4).

Apart from single rare-earth composition $Re$NiO$_3$, the perovskite nickelates with binary and triple rare-earth compositions were also investigated to achieve a more continuous regulation of the $T_{R\text{-}MIT}$ and its resultant delta-temperatural transports. It was previously known that for regulating the MIT of rare-earth nickelates, the $T_{MIT}$ for $Re_x$Re'$_{1-x}$NiO$_3$ is capable to be linearly adjusted between the $T_{MIT}$ of $Re$NiO$_3$ and $Re$'NiO$_3$, via varying their relative composition of $x$ [8,26]. Figure 4a shows the $R$-$T$ relation in the low temperature range for the multiple rare-earth composition perovskite nickelates, while their $T_{R\text{-}MIT}$ are more clearly compared in Figure 4b with the single rare-earth composition $Re$NiO$_3$. In contrast to MIT, it is interesting to note that as observed $T_{R\text{-}MIT:\,Re,Re'}$ for $Re_x$Re'$_{1-x}$NiO$_3$ is always below the calculated one from the composition weighted average of $T_{R\text{-}MIT}$ from $Re$NiO$_3$ and $Re$'NiO$_3$, as [$xT_{R\text{-}MIT:Re}+(1-x)\ T_{R\text{-}MIT:Re'}$]. As several representative examples demonstrated in Figure 4b, the Sm$_{3/4}$Tm$_{1/4}$NiO$_3$ exhibits a lower $T_{R\text{-}MIT}$ compared to SmNiO$_3$ and TmNiO$_3$; the Sm$_{3/4}$Eu$_{1/4}$NiO$_3$ exhibits a lower $T_{R\text{-}MIT}$ compared to SmNiO$_3$ and EuNiO$_3$; and the Sm$_{3/4}$Tm$_{1/4}$NiO$_3$ exhibits a lower $T_{R\text{-}MIT}$ compared to SmNiO$_3$ and TmNiO$_3$.

These results further confirm our understanding that the R-MIT is the entropy dominated transportation behavior that differs to the MIT in conventional Mott-Hubbard systems as dominantly driven by the orbital Coulomb energy [8]. It is worth noticing that the difference in rare-earth composition within $Re$NiO$_3$ is expected to vary the orbital configurations in every freedom within the real space. Therefore, the elevation in $S_{Orbit}$ for introducing an additional rare-earth composition is expected to be much larger compared to the enhancement in the compositional entropy that is simply calculated as $\Delta S_{EPC} = -k_B[x\log(x) + (1-x)\log(\frac{1-x}{1-N})]$ [16].

As illustrated in Figure 4c, the elevation in $S_{Orbit}$ via multiple compositional $Re$ is expected to enhance the negative gain in the entropy contribution to $\Delta G_{Orbit}$ within the insulating phase of $Re_x$Re'$_{1-x}$NiO$_3$, resulting in the reduction of $T_{R\text{-MIT}}$. The $R_{T\text{-Delta}}/R_{300K}$ and $T_{Delta}$-FWHM for multiple rare-earth composition perovskite nickelates are shown in Figure S5, where a reducing sharpness in delta-temperatural transport is also observed with the elevation of $T_{R\text{-MIT}}$. Although the relationship of $T_{R\text{-MIT}}$ with the multiple-$Re$ compositions is more complex than the one of $T_{MIT}$, a continuous regulation in the magnitude of $T_{R\text{-MIT}}$ can be still achieved. As achieved delta-temperatural transport is expected to open up a new gate for exploring new applications, such as more conveniently locking/excluding the working conditions of electric devices and circuits within a narrow window of temperature.

In summery, an additional electronic phase transition at a lower temperature and with a reverse-temperature dependent transportation behavior compared to metal to insulator transition is discovered within metastable perovskite family of correlated $Re$NiO$_3$ (the size of $Re$ <Nd). We name such transition as reverse-metal to insulator transition (R-MIT) and achieve broad regulations in its transition point ($T_{R\text{-MIT}}$) via adjusting the rare-earth composition or imparting interfacial strains. The R-MIT is expected to be driven by the entropy contribution to the orbital free energy that determines the relative stability between the metallic and insulating phase, and differs to the Coulomb energy driven MIT. This is evidenced by the lower $T_{R\text{-MIT:Re,Re'}}$ observed in multiple rare-earth composition perovskites nickelates (i.e., $Re_x$Re'$_{1-x}$NiO$_3$) compared to the one from the composition weighted average of $T_{R\text{-MIT}}$ from $Re$NiO$_3$ and $Re'$NiO$_3$ as [$xT_{R\text{-MIT:Re}}+(1-x)$ $T_{R\text{-MIT:Re'}}$]. From the fundamental aspect, the thermodynamically stable temperature range for insulating phase of $Re$NiO$_3$ is redefined as between of $T_{R\text{-MIT}}$ and $T_{MIT}$, while further descending temperature below $T_{R\text{-MIT}}$ triggers orbital re-configuration towards a a new conductive phase with high orbital entropy. From the aspect of application, a distinguished character of delta-temperatural transports is achieved via combining the R-MIT and afterwards NTCR thermistor transportations of $Re$NiO$_3$ by elevating the temperature across $T_{R\text{-MIT}}$. It results in a significant enhancement in the electronic resistivity within a narrow window of temperature, and is expected to result in new applications such as locking the working temperatures range for electric devices catering for the demand in the fast developed automatic transmission and artificial intelligence.

**Acknowledgments**
This work was supported by National Natural Science Foundation of China (No. 51602022 and No. 61674013). We also acknowledge the technical support or discussions from Prof. Akira Toriumi from the University of Tokyo and Prof. Lidong Chen from Shanghai Institute of Ceramics Chinese Academy of Sciences.

**Competing interests**
We declare no competing financial interest.

**Additional information:** Supplementary Information is available for this manuscript.

**Correspondences:** Correspondence should be addressed: Prof. Jikun Chen (jikunchen@ustb.edu.cn).

**Author contributions:** JC proposed the original idea, planed for the work, performed partially the experiment (i.e., sample growth, structure and transport characterizations), analysed the data, and write the manuscript; HH contributed in the sample growth and transport characterizations; HD and BG contributed to the TEM experiment; JW contributed for the EXAFS experiment; TY provides constructive discussions; NC and YJ provide constructive support from the aspects of sample growth and characterization, respectively.

**Figures and captions**

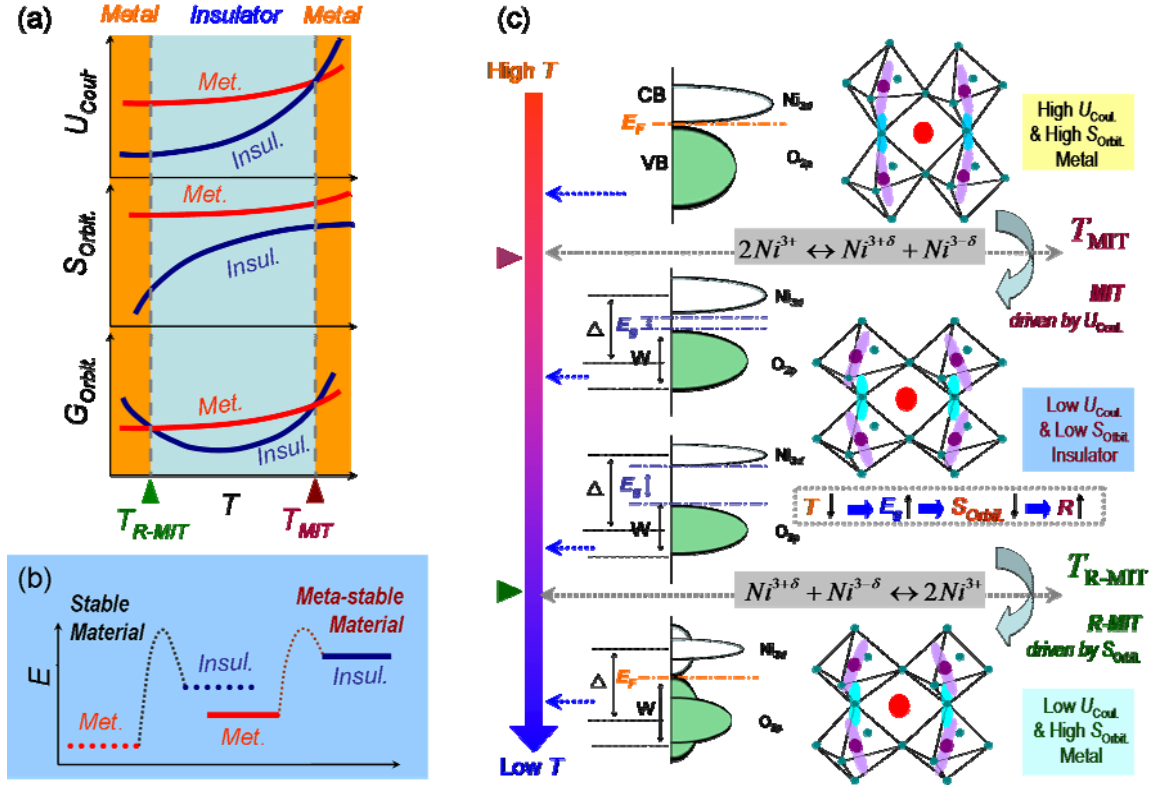

**Figure 1.** (a) The temperature dependence of the Coulomb repulsion energy ($U_{Coul.}$), orbital entropy ($S_{Orbit.}$) and free energy ($G_{Orbit.}$) as illustrated for the insulating phase and metallic phase of $Re$NiO$_3$. By reducing the temperature across $T_{MIT}$, the $U_{Coul.}$ reduces more significantly for the insulating phase compared to the metallic one that triggers the conventional metal to insulator transitions (MIT). Afterwards, the band gap gradually opens by further reducing the temperature, and the ordering in orbital charge of Ni$^{2+}$ and Ni$^{4+}$ is improved to descend $S_{Orbit.}$, which elevates the $G_{Orbit.}$ of the insulating phase compared to the metallic one. When reaching $T_{R-MIT}$, the lower $U_{Coul.}$ of the insulating phase compared to the metallic phase was offset by the reduction in $S_{Orbit.}$ owning to the high orbital ordering, and this thermodynamically triggers the orbital transformation to reduce their ordering. As a result, a reversed metal to insulating transition (R-MIT) occurs to transform $Re$NiO$_3$ from the insulating phase at to metallic phase by further reducing the temperature. (b) The material metastability is expected to reduce the energy barrier to overcome when triggering the R-MIT. (c) Orbital and structural transformations when the $Re$NiO$_3$ transforms from the metal to insulator (MIT) and further to metal (R-MIT) with a reducing temperature.

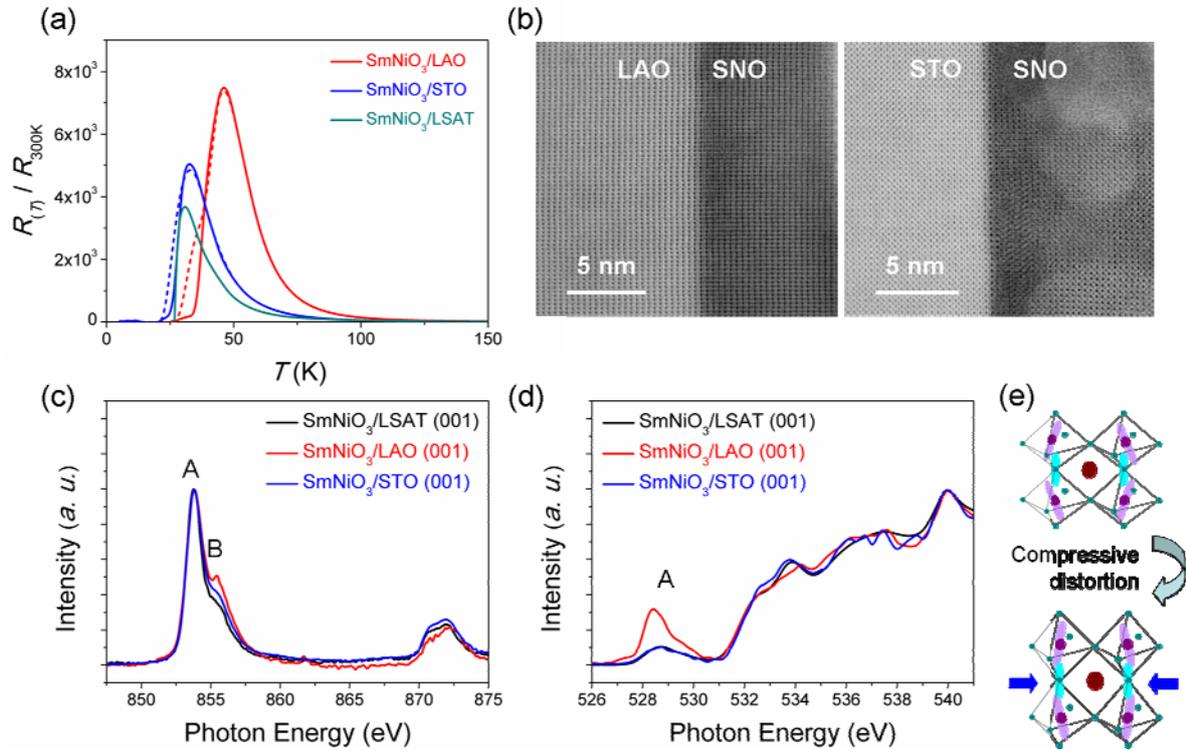

**Figure 2.** **(a)** Temperature-dependence of the material resistivity ($R$-$T$) for SmNiO$_3$ (SNO) on the LaAlO$_3$ (LAO), SrTiO$_3$ (STO) and (LaAlO$_3$)$_{0.3}$(Sr$_2$AlTaO$_6$)$_{0.7}$ (LSAT) substrates with a (001) orientation. The solid lines were measured via heating up, while the dash lines were measured via cooling down. **(b)** Representative cross-section morphologies for SmNiO$_3$/LAO and SmNiO$_3$/STO form the high-angle annular dark-field (HAADF). **(c)-(e)** Near edge X-ray absorption fine structure (NEXAFS) analysis of **(c)** Ni-$L_3$ edge and **(d)** O-$K$ edge of SmNiO$_3$ at various states of interfacial strains, while the respective variation in the NiO$_6$ octahedron under bi-axial compressive distortion is illustrated in **(e)**.

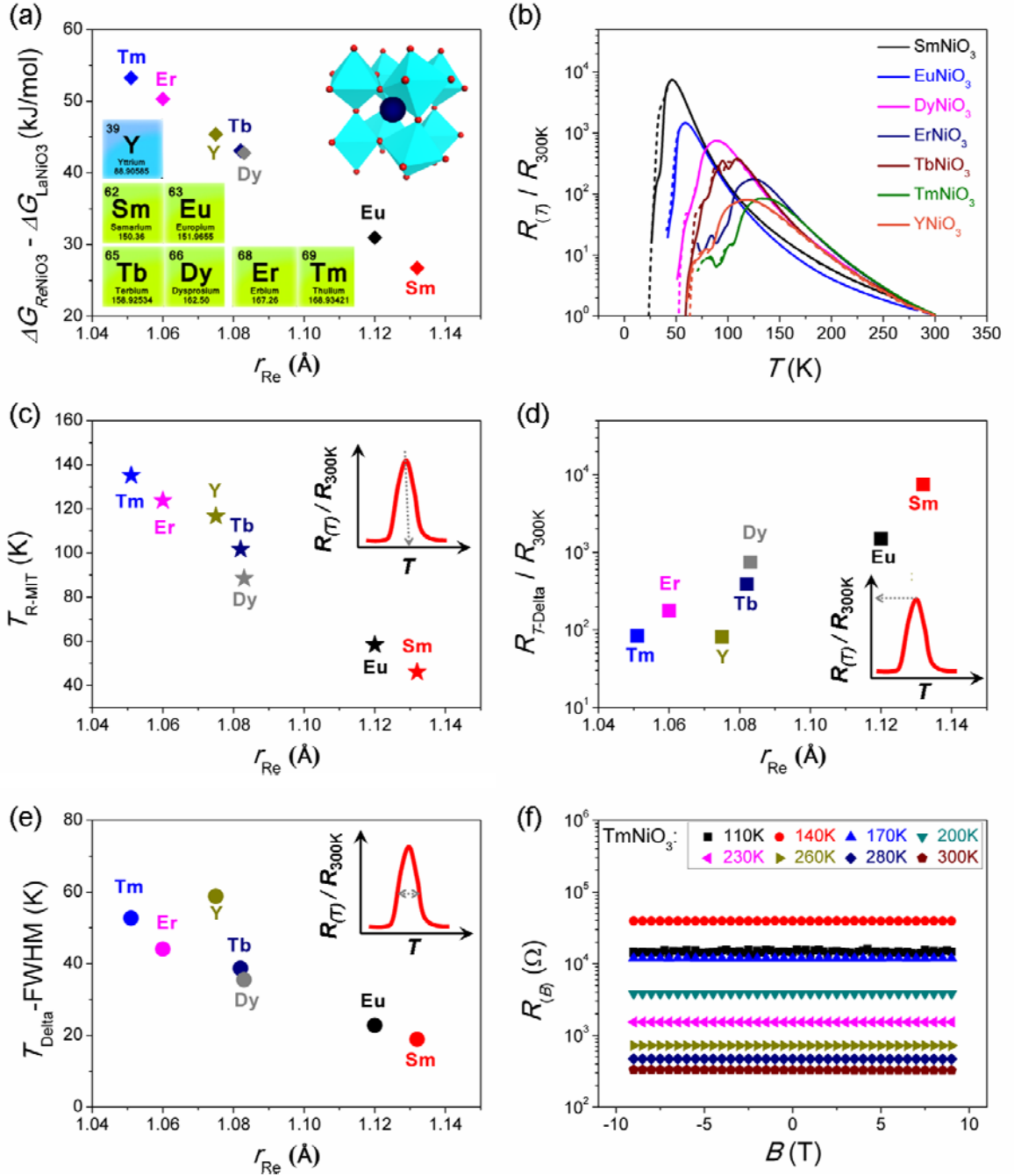

**Figure 3.** (a) The elevation in material meta-stability for $Re$NiO$_3$, compared to the thermodynamically stable LaNiO$_3$, with an reducing the size of the rare-earth elements. (b) Temperature-dependence of the material resistivity ($R$-$T$) for single rare-earth composition $Re$NiO$_3$ grown on the LaAlO$_3$ (001) substrate. The solid lines were measured via heating up, while the dash lines were measured via cooling down. (c) The reverse metal to insulator transition temperature ($T_{\text{R-MIT}}$), (d) the maximum resistivity at $T_{\text{R-MIT}}$ compared to the one at 300 K, and (e) the full widths half maximum of its resultant delta-shaped temperature dependence in resistivity ($T_{\text{Delta}}$-FWHM) summarized from the $R$-$T$ of $Re$NiO$_3$. (f) Resistance of TmNiO$_3$ measured as a function of the imparted external magnetic fields ($B$) at various temperatures.

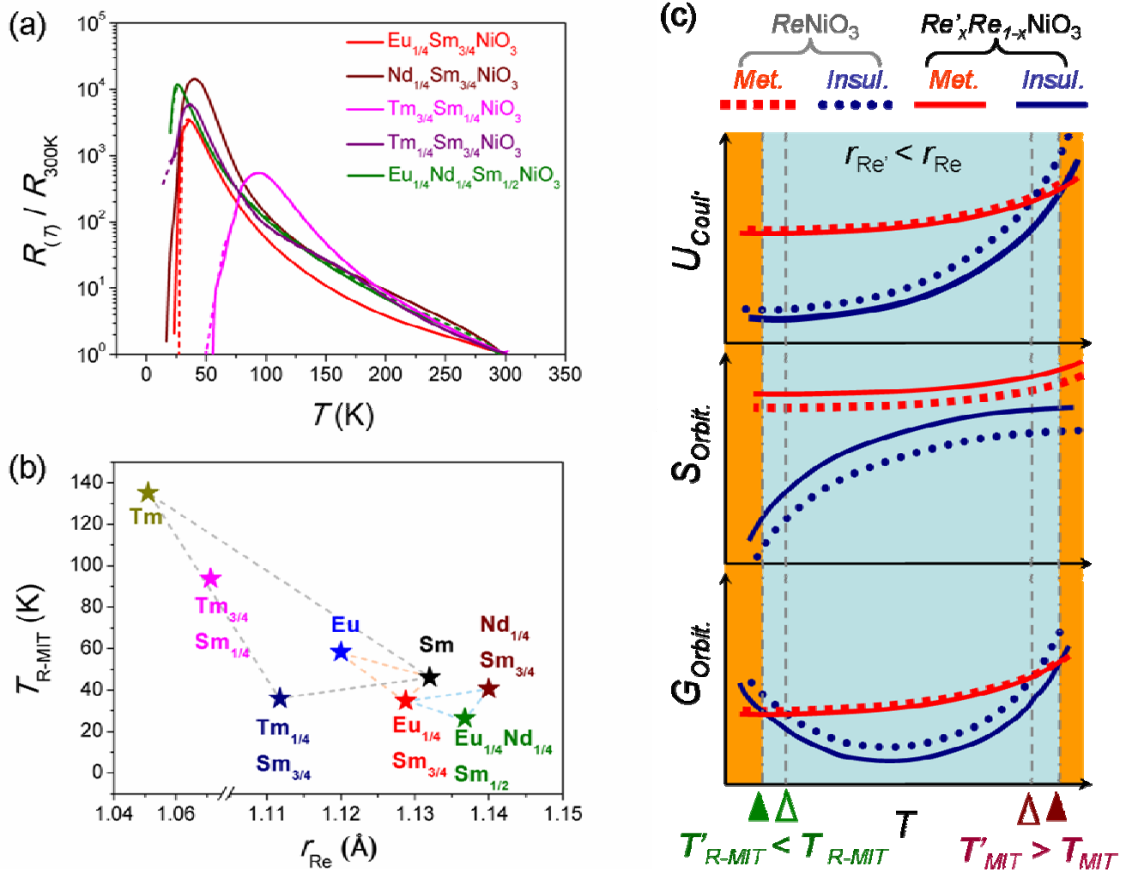

**Figure 4.** **(a)** Temperature-dependence of the material resistivity ($R$-$T$) for multiple rare-earth compositional $Re$NiO$_3$ grown on the LaAlO$_3$ (001) substrate. The solid lines were measured via heating up, while the dash lines were measured via cooling down. **(b)** The reverse metal to insulator transition temperature ($T_{\text{R-MIT}}$) summarized from the $R$-$T$ of multiple compositional $Re$NiO$_3$. **(c)** Illustrating the variations in the Coulomb repulsion energy ($U_{\text{Coul.}}$), orbital entropy ($S_{\text{Orbit.}}$) and free energy ($G_{\text{Orbit.}}$) when substitute the rare-earth composition in $Re$NiO$_3$ by a smaller rare-earth element ($Re'$). This is previously known to more distort the NiO$_6$ octahedron that opens the band gap and elevate $T_{\text{MIT}}$. Unlike the MIT mainly driven by $U_{\text{Coul.}}$, the R-MIT is expected to be triggered by the contribution from the descending $S_{\text{Orbit}}$ to $G_{\text{Orbit}}$ with temperature that offsets the variation in $U_{\text{Coul.}}$ between the insulating and metallic phases. The multiple rare-earth composition within $Re$NiO$_3$ is expected to not only enhance the compositional entropy but also largely enrich the orbital configuration complexity that elevates $S_{\text{Orbit.}}$, and this further results in the reduction in $T_{\text{R-MIT}}$.

**References**

[1] Lu, N; Zhang, P.; Zhang, Q.; Qiao, R.; He, Q.; Li, H. B.; Wang, Y.; Guo, J.; Zhang, D.; Duan, Z.; Li, Z.; Wang, M.; Yang, S.; Yan, M.; Arenholz, E.; Zhou, S.; Yang, W.; Gu, L.; Nan, C. W.; Wu, J.; Tokura Y.; Yu, P. Electric-field control of tri-state phase transformation with a selective dual-ion switch. *Nature* **2017**, 546, 124

[2] Kung, H. H.; Baumbach, R. E.; Bauer, E. D.; Thorsmølle,V. K.; Zhang, W. L.; Haule, K.; Mydosh, J. A.; Blumberg, G. Chirality density wave of the "hidden order" phase in $URu_2Si_2$. *Science* **2015**, 347, 1339-1342

[3] Li, L. J.; O'Farrell, E. C. T.; Loh, K. P.; Eda, G.; Özyilmaz, B.; Castro Neto, A. H., Controlling many-body states by the electric-field effect in a two-dimensional material. *Nature* **2016**, 529, 185

[4] Zhang, Z.; Schwanz, D.; Narayanan, B.; Kotiuga, M.; Dura, J. A.; Cherukara, M.; Zhou, H.; Freeland, J. W.; Li, J.; Sutarto, R.; He, F.; Wu, C.; Zhu, J.; Sun, Y.; Ramadoss, K.; Nonnenmann, S. S.; Yu, N.; Comin, R.; Rabe, K. M.; Sankaranarayanan, S. K. R. S.; Ramanathan, S. Rerovskite nickelates as electric-field sensor in salt water. *Nature* **2018**, 553, 68

[5] Zhou, Y.; Guan, X.; Zhou, H.; Ramadoss, K.; Adam, S.; Liu, H.; Lee, S.; Shi, J.; Tsuchiya, M.; Fong D. D.; Ramanathan, S. Strongly correlated perovskite fuel cells. *Nature* **2016**, 534, 231

[6] Shi, J.; Zhou, Y.; Ramanathan, S. Colossal resistance switching and band gap modulation in a perovskite nickelate by electron doping. *Nat. Commun*. **2014**, 5, 4860

[7] Goodenough, J. B. The two components of crystallographic transition in $VO_2$. *J. Solid State Chem.* **1971**, 3, 490–500

[8] Catalan, G.; Progress in perovskite nickelate research. *Phase Transitions* **2008**, 81, 729

[9] K. Shibuya, M. Kawasaki, and Y. Tokura, Metal-insulator transition in epitaxial $V_{1-x}W_xO_2$ (0< x<0.33) thin films. *Appl. Phys. Lett.* **2010,** *96*, 022102

[10] Guo, F.; Chen, S.; Chen, Z.; Luo, H.; Gao, Y.; Przybilla, T.; Spiecker, E.; Osvet, A.; Forberich, K.; Brabec, C. J., Printed smart photovoltaic window integrated with an energy-saving thermochromic layer. *Adv. Optical Mater.* **2015**, 3, 1524–1529

[11] Chen, J.; Hu, H.; Wang, J.; Yajima, T.; Ge, B.; Ke, X.; Dong, H.; Jiang, Y.; Chen, N., Overcoming synthetic metastabilities and revealing metal-to-insulator transition & thermistor bi-functionalities for d-band correlation perovskite nickelates. *Mater. Horizons* **2019**, DOI: 10.1039/c9mh00008a

[12] Shi, J.; Ha, S. D.; Zhou, Y.; Schoofs, F.; Ramanathan, S. A correlated nickelate synaptic

transistor. *Nat. Commun.* **2013**, 4, 2676

[13] Martens, K.; Jeong, J. W.; Aetukuri, N.; Rettner, C.; Shukla, N.; Freeman, E.; Esfahani, D. N.; Peeters, F. M.; Topuria, T.; Rice, P. M.; Volodin, A.; Douhard, B.; Vandervorst, W.; Samant, M. G.; Datta, S.; Parkin, S. S. P., Field effect and strongly localized carriers in the metal-insulator transition material $VO_2$. *Phys. Rev. Lett.* **2015**, 115, 196401

[14] Yajima, T.; Nishimura, T.; Toriumi, A. Positive-bias gate-controlled metal–insulator transition in ultrathin $VO_2$ channels with $TiO_2$ gate dielectrics. *Nat. Commun.* **2015**, 6, 10104

[15] Zuo, F.; Panda, P.; Kotiuga, M.; Li, J.; Kang, M.; Mazzoli, C.; Zhou, H.; Barbour, A.; Wilkins, S.; Narayanan, B.; Cherukara, M.; Zhang, Z.; Sankaranarayanan, S. K. R. S.; Comin, R.; Rabe, K. M.; Roy K.; Ramanathan S. Habituation based synaptic plasticity and organismic learning in a quantum perovskite. *Nat. Commun.* **2017**, 8, 240

[16] Budai, J. D.; Hong, J.; Manley, M. E.; Specht, E. D.; Li, C. W.; Tischler, J. Z.; Abernathy, D. L.; Said, A. H.; Leu, B. M.; Boatner, L. A.; McQueeney R. J.; Delaire, O., Metallization of vanadium dioxide driven by large phonon entropy. *Nature* **2014**, 515, 535

[17] Rost, C. M.; Sachet, E.; Borman, T.; Moballegh, A.; Dickey, E. C.; Hou, D.; Jones, J. L.; Curtarolo, S.; Maria, J. P., Entropy-stabilized oxides. *Nat. Commun.* **2015**, 6, 8485

[18] I. I. Mazin, D. I. Khomskii, R. Lengsdorf, J. A. Alonso, W. G. Marshall, R. M. Ibberson, A. Podlesnyak, M. J. Martínez-Lope, and M. M. Abd-Elmeguid, Charge Ordering as Alternative to Jahn-Teller Distortion. *Phys. Rev. Lett.* **2007**, *98*, 176406

[19] J. S. Zhou and J. B. Goodenough, Chemical bonding and electronic structure of $R$NiO$_3$ ($R$=rare earth), *Phys. Rev. B* **2004**, *69*, 153105

[20] J. A. Alonso, J. L. García-Muñoz, M. T. Fernández-Dı́az, M. A. G. Aranda, M. J. Martínez-Lope, and M. T. Casais, Charge Disproportionation in $R$NiO3 Perovskites: Simultaneous metal-insulator and structural transition in YNiO$_3$. *Phys. Rev. Lett.* **1999,** *82*, 3871

[21] K. Kleiner, J. Melke, M. Merz, P. Jakes, P. Nage, S. Schuppler, V. Liebau, H. Ehrenberg, Unraveling the degradation process of $LiNi_{0.8}Co_{0.15}Al_{0.05}O_2$ electrodes in commercial lithium ion batteries by electronic structure investigations, *ACS Appl. Mater. Interfaces* **2015**, *7*, 19589.

[22] L. A. Montoro, M. Abbate, J. M. Rosolen, Electronic structure of transition metal Ions in deintercalated and reintercalated $LiCo_{0.5}Ni_{0.5}O_2$. *J. Electrochem. Soc.* **2000**, *147*, 1651−1657.

[23] F. Reinert, P. Steiner, S. Hüfner, H. Schmitt, J. Fink, M. Knupfer, P. Sandl, E. Bertel, Electron and hole doping in NiO. *Z. Phys. B: Condens. Matter.* **1995**, *97*, 83−93.

[24] Chen, J.; Mao, W.; Ge, B.; Wang, J.; Ke, X.; Wang, V.; Wang, Y.; Döbeli, M.; Geng, W.;

Matsuzaki, H.; Shi, J.; Jiang, Y., Revealing the role of lattice distortions in the hydrogen-induced metal-insulator transition of SmNiO$_3$. *Nat. Commun.* **2019**, 10, 694

[25] Conchon, F.; Boulle, A.; Guinebretière, R. Effect of tensile and compressive strains on the transport properties of SmNiO$_3$ layers epitaxially grown on (001) SrTiO$_3$ and LaAlO$_3$ substrates. *Appl. Phys. Lett.* **2007**, 91, 192110

[26] Ambrosini, A.; Hamet, J. F. Sm$_x$Nd$_{1-x}$NiO$_3$ thin-film solid solutions with tunable metal–insulator transition synthesized by alternate-target pulsed-laser deposition. *Appl. Phys. Lett.* **2003,** 82, 727

[27] Escote, M. T.; da Silva, A. M. L.; Matos, J. R.; Jardim, R. F. General properties of polycrystalline *Ln*NiO$_3$ (Ln = Pr, Nd, Sm) compounds prepared through different precursors. *Journal of Solid State Chemistry* **2000**, 151, 298-307

# Supporting Information

# Entropy driven reverse-metal-to-insulator transition and delta-temperatural transports in metastable perovskites of correlated rare-earth nickelate

*Jikun Chen[1], Haiyang Hu[1], Takeaki Yajima[2], Jiaou Wang[3], Binghui Ge[4], Hongliang Dong[5], Yong Jiang[1],*

*Nuofu Chen[6], et al*

[1]Beijing Advanced Innovation Center for Materials Genome Engineering, School of Materials Science and Engineering, University of Science and Technology Beijing, Beijing 100083, China

[2]School of Engineering, The University of Tokyo, 2-11-16 Yayoi, Bunkyo-ku, Tokyo 113-0032, Japan

[3]Beijing Synchrotron Radiation Facility, Institute of High Energy Physics, Chinese Academy of Sciences, Beijing 100049, China

[4]Institute of Physical Science and Information Technology, Anhui University, 230601, Heifei, Anhui, China
Cleveland, Ohio 44106, United States

[5]Center for High Pressure Science and Technology Advanced Research, Shanghai 201203, China

[6]School of Renewable Energy, North China Electric Power University, Beijing 102206, China

Correspondence and request for materials: Prof. Jikun Chen (jikunchen@ustb.edu.cn)

**Section A: Experimental details**

**Sample growth:** The chemical growth of $Re$NiO$_3$ includes the following three steps: (1) The chemical precursors of $Re$(NO$_3$)$_3$ and Ni(CH$_3$COOH)$_2$ were mixed at the nominal stoichiometry within ethylene glycol monomethyl ether (EGME) under ultrasonic. (2) As made chemical solutions were spin coated onto single crystalline substrates of LaAlO$_3$, SrTiO$_3$ and (LaAlO$_3$)$_{0.3}$(Sr$_2$AlTaO$_6$)$_{0.7}$, with the (001) orientation and one side polished, followed by baking at 175 °C to evaporate the EGME. (3) To achieve the crystallization as perovskite $Re$NiO$_3$, the samples were annealed at 800 °C within oxygen pressure of 15-20 MPa for 3 hours. The pulsed laser deposition of SmNiO$_3$ was performed by laser ablating of a ceramic target with nominal compositions in 20 Pa O$_2$ pressure within a vacuum chamber. During the deposition, the temperature of the SrTiO$_3$ substrate was kept as 650 °C. After the deposition process, as obtained sample was annealed within oxygen pressure of 15 MPa for 3 hours.

**Characterizations:** A commercialized CTA-system was used to measure the resistivity of as-grown thin films in high temperature range within 300 K- 550 K, while using a PPMS system from Quantum Design was used to characterize their resistance in the low temperature range within 5 K – 400 K and under external magnetic fields. The cross-plane and in-plane information of as-grown films compared to the substrates were probed by using reciprocal space mapping (RSM). The [114] reciprocal space vectors of both the film and substrate are projected at the [110] and [001] reciprocal vectors for the in-plane and cross-plane direction, respectively. To characterize the cross-plane morphology of as-grown thin films, the high-angle annular dark-field (HAADF) and annular bright-field (ABF) scanning transmission electron microscopy (STEM) was performed on JEM-ARM 200F TEM operated at 200 kV with a cold field emission gun and aberration correctors for both probe-forming and imaging lenses. To characterize the electrical orbital structures, we performed the near edge X-ray absorption fine structure (NEXAFS) at Beijing Synchrotron Radiation Facility, Institute of High Energy Physics, Chinese Academy of Sciences, Beijing 100049, China.

**Section B: Additional results**

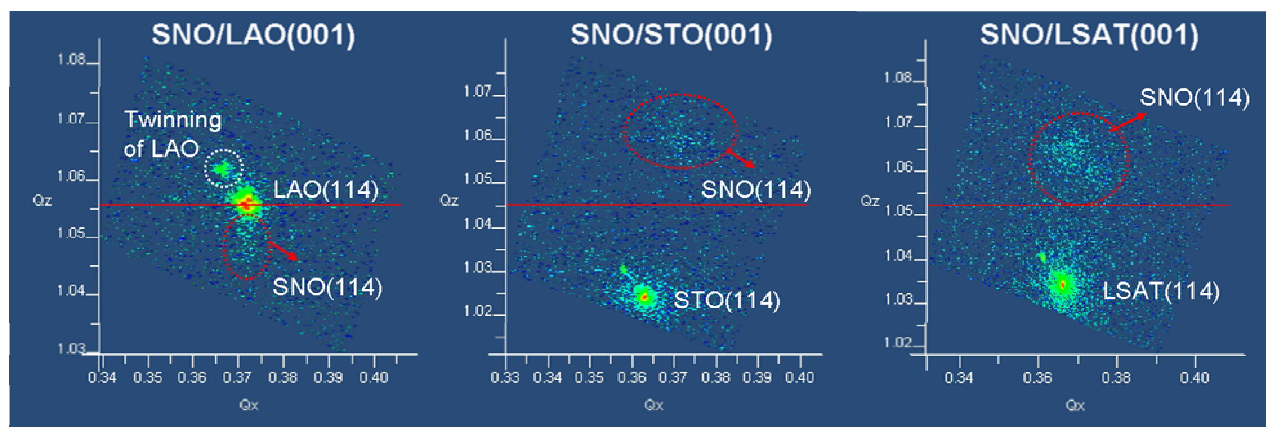

**Figure S1.** The X-ray reciprocal space mapping of SmNiO$_3$ (SNO) grown on single crystalline perovskite structured substrates, such as LaAlO$_3$ (LAO), SrTiO$_3$ (STO) and (LaAlO$_3$)$_{0.3}$(Sr$_2$AlTaO$_6$)$_{0.7}$ (LSAT) with (001) orientation via the chemical approach reported previously. The reciprocal vector of (114) is used to probe the diffraction patterns from both the thin film and the substrate.

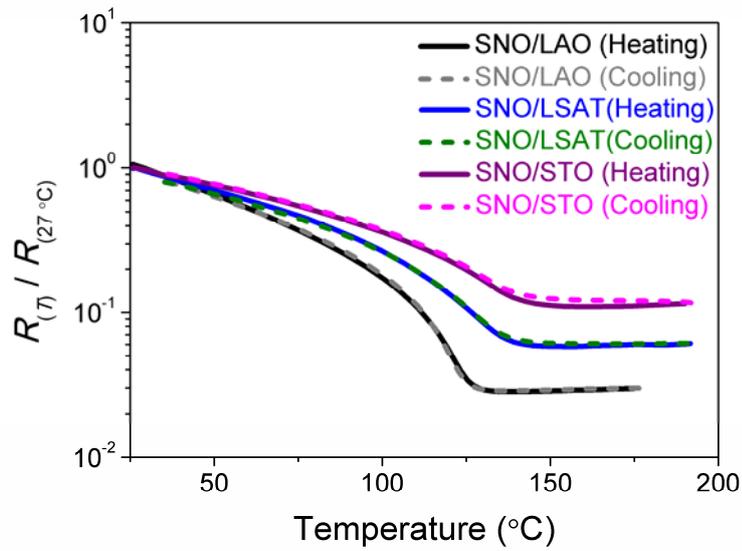

**Figure S2.** The temperature dependence in resistivity for $SmNiO_3/LaAlO_3$, $SmNiO_3/SrTiO_3$, and $SmNiO_3/(La,Sr)(Al,Ta)O_3$, measured via heating up (solid lines) and cooling down (dash lines). A lower metal to insulator transition temperature ($T_{MIT}$) is observed for the bi-axial distorted $SmNiO_3/(La,Sr)(Al,Ta)O_3$, compared to the partially tensile strain relaxed $SmNiO_3/SrTiO_3$, and $SmNiO_3/(La,Sr)(Al,Ta)O_3$. These observations are in agreement to the previous reports on strain distorted $SmNiO_3$.

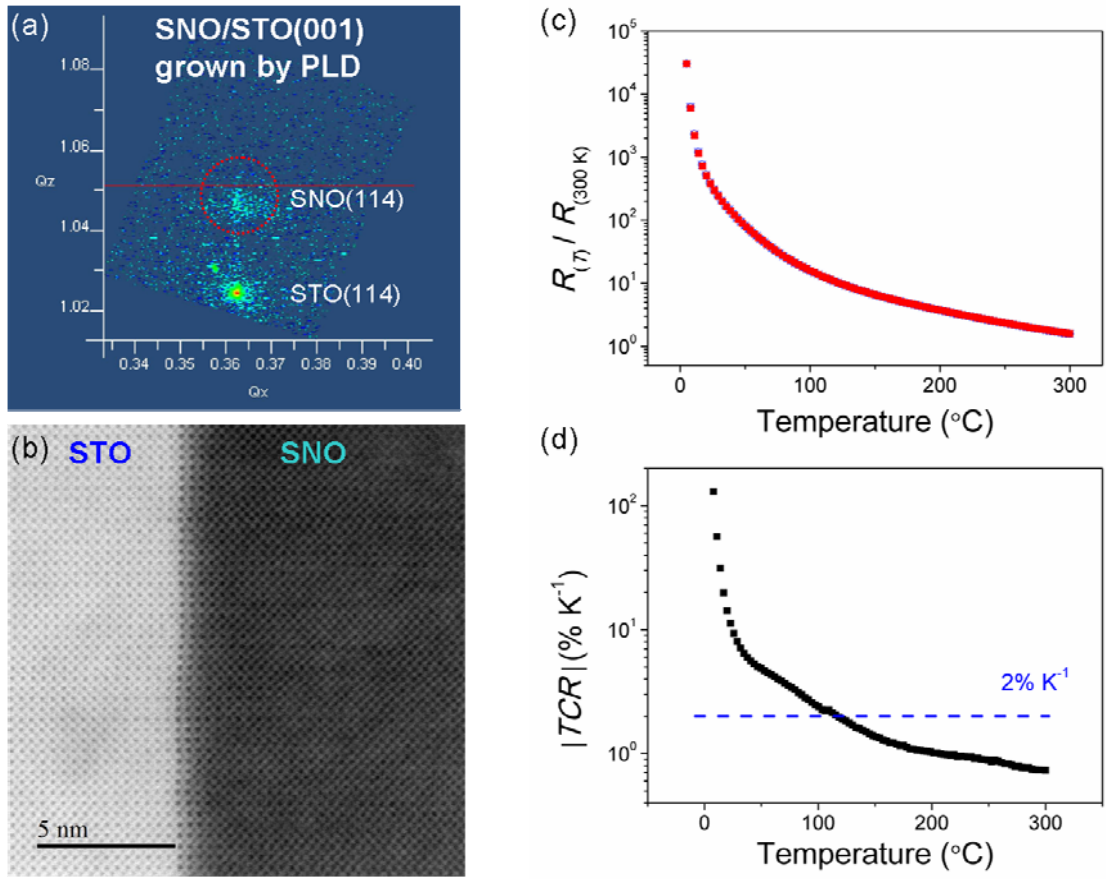

**Figure S3. (a)** The X-ray reciprocal space mapping of SmNiO$_3$ (SNO) grown on single crystalline SrTiO$_3$ (STO) with (001) orientation via pulsed laser deposition. The reciprocal vector of (114) is used to probe the diffraction patterns from both the thin film and the substrate. **(b)** The interfacial morphology of as-grown SmNiO$_3$/SrTiO$_3$ from the high-angle annular dark-field (HAADF) images. **(c)** Temperature dependence of the resistivity and **(d)** Temperature coefficients of resistance (TCR) for as-grown SmNiO$_3$/SrTiO$_3$.

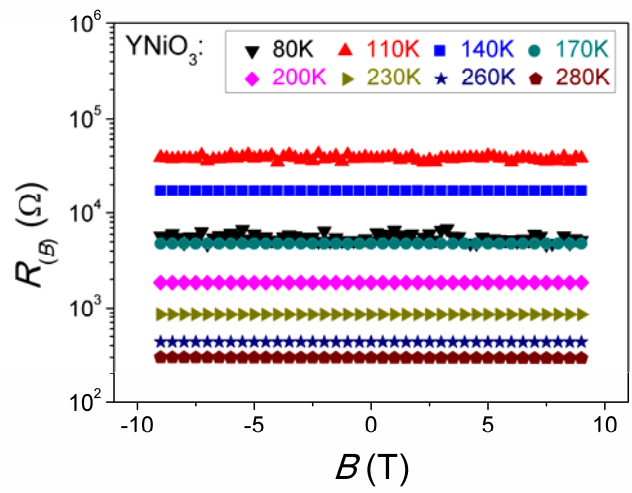

**Figure S4.** Resistance of YNiO$_3$ measured as a function of the imparted external magnetic fields (*B*) at various temperatures.

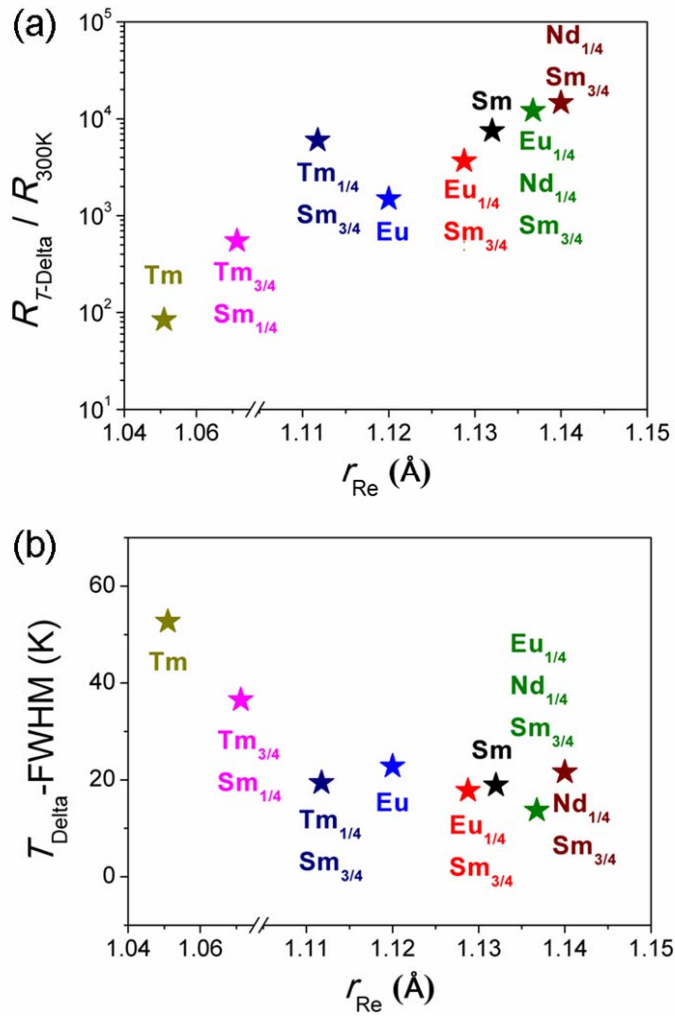

**Figure S5.** (a) The maximum resistivity at $T_{\text{R-MIT}}$ compared to the one at 300 K, and (b) the full widths half maximum of its resultant delta-shaped temperature dependence in resistivity ($T_{\text{Delta}}$-FWHM) summarized from the $R$-$T$ of multiple rare-earth composition perovskite nickelates shown in Figure 4a.